\pdfoutput=1
\documentclass[preprint,twocolumn,tightenlines,10pt,prl]{revtex4}%
\usepackage{amsfonts}
\usepackage{amsmath}
\usepackage{amssymb}
\usepackage{graphicx}%
\providecommand{\U}[1]{\protect\rule{.1in}{.1in}}

\begin{document}
\title{Hexagonal spiral growth in the absence of a substrate}
\author{L. L. A. Adams*}
\affiliation{The James Franck Institute, The University of Chicago, 929 East 57th St.,
Chicago, IL 60637}


\begin{abstract}
Experiments on the formation of spiraling hexagons (350 - 1000 nm in width)
from a solution of nanoparticles are presented. Transmission electron
microscopy images of the reaction products of chemically synthesized cadmium
nanocrystals indicate that the birth of the hexagons proceeds without
assistance from static screw or edge dislocatons, that is, they spiral without
constraints provided by an underlying substrate. Instead, the apparent growth
mechanism relies on what we believe is a dynamical dislocation identified as a
dense aggregate of small nanocrystals that straddles the spiraling hexagon at
the crystal surface. This nanocrystal bundle, which we term the
\textquotedblleft feeder\textquotedblright, also appears to release
nanocrystals into the spiral during the growth process. \TeX{} .

\end{abstract}
\volumeyear{year}
\volumenumber{number}
\issuenumber{number}
\received[Received text]{date}

\revised[Revised text]{}

\startpage{1}
\endpage{102}
\maketitle

Crystals and their faceted faces are beautiful and remarkable especially when
considering how their atomic constituents structurally organize. How do atoms
in crystal growth processes "know" when to turn left or right, or stop growing
in a particular direction to pursue a new direction? When is a crystal a
crystal and no longer a fluid?

The answers to these questions are difficult to address experimentally because
of the challenges of catching atoms in the act of crystallizing.\cite{Yau}
However there are clues that allow one to confirm existing theories concerning
the nature of the underlying growth mechanism.\ Take, for example, the case of
screw dislocations. In 1949 Frank \cite{Frank1} had the insight that crystals
emerge from the existence of screw dislocations that allow them to grow from
stepped surfaces. This controversial claim was based on what Frank noticed as
large discrepancies between experimental growth rates and the inability of
two-dimensional nucleation theory to account for them. After much skeptism
from the crystal field community \cite{Cahn}, Frank's theory was confirmed by
the appearance of spiraling constructs on the surface of carborundum when
decorated with minute amounts of silver and looking at the resulting
refractory patterns.\cite{verma} Thus, the concept of screw dislocations
manifested by spiraling crystals was not only verified but also became the
impetus to a new way of thinking about crystal growth mechanisms. And today
this active area of research continues to find growth spirals in materials
such as thin films of high temperature superconductors \cite{Gerber}, metals
\cite{Redinger} , gallium nitride \cite{Strunk} and organic materials such as
pentacene \cite{Ruiz} to name a few. Screw dislocations may even be
responsible for supersolidity in $^{4}$He \cite{Chan} as demonstrated by
several theorists using Monte Carlo simulations. \cite{Troyer}

If the crystal growth process advances without a substrate \cite{vinokur} is
it still possible to form faceted growth spirals? In this Letter we report
experimental evidence which indicates that a substrate with screw (or edge)
dislocations is not essential for hexagonal spiral crystals, at least not in
two dimensions (2D). Surprisingly stable 2D\ spiral crystals can be produced
by chemical synthesis techniques employed in the reduction of metallic salts.
As will be addressed in this paper, these spirals appear to be governed by a
dense aggregate of small particles believed to be two dimensional nuclei. This
dense cluster intersects at the surface of the hexagonal spiral and possibly
serves as a dynamical dislocation moving along the periphery of the hexagon
and discharging particles into the spiral. Hereafter, we refer to this cluster
as the "feeder". But before discussing the growth of 2D spiral crystals, the
overall process of forming 2D hexagons is described.

Two dimensional cadmium hexagons were synthesized by reduction of cadmium
acetate in a three neck flask with oleic acid and trioctylphosphine serving as
ligands in the presence of a solvent (octyl ether or octadecene). Besides
serving as a ligand, oleic acid, as the dominate player in these reactions,
also helps reduce cadmium acetate, slow down the reaction, and control the
shape of the resulting nanocrystals. Large concentrations of oleic acid, a
very long chain organic molecule, prevent the formation of hexagons all
together indicating that the competition between cadmium's inclination to form
hexagons is blocked by oleic acid's polar and nonpolar composition. Since
cadmium nanocrystals produced in these reactions are extremely nonpolar, its
nonpolarity favors the large nonpolar portion of oleic acid while it is
strongly repulsed by oleic acid's polar end. Finally it should be emphasized
that cadmium is a carcinogen, so careful handling must be carried out. (A
detailed description of the synthesis will be published elsewhere.)

With the proper selection of reaction time, temperature, stirring speed and
relative concentration between the precusor and ligands, hexagons and
spiraling hexagons can be synthesized. \ Figure 1 shows the transmission
electron microscopy (TEM) image for the case of transparent 2D hexagons
containing both symmetric and antisymmetric shapes. These TEM images are
prepared by pipeting a single drop of the toluene nanocrystal solution onto a
TEM\ grid and imaging at 300 keV using a FEI\ Tecnai F30 instrument. Due to
the thinness of these nanocrystals, the large surface area to volume ratio,
and the fact that these nanocrystals are not isolated from one another makes
them susceptible to bending forces. Evidence of this is manifested in the bend
contours that are seen from Bragg diffraction effects shown as broad dark and
light bands extending outwards from a central point. It is also possible to
observe Moire patterns (shown in the inset) when two different crystalline
hexagons lie on top of one another. In this inset, the top hexagon is
partially folded over which creates bending contours discernible in the
hexagon lying beneath.

Bulk cadmium, itself, has some interesting properties which might explain a
few of the features presented here. Cadmium, a hexagonally closed packed (hcp)
crystal, naturally likes to form hexagons as shown in Fig. 1; in addition, of
all the hcp elements, it has the largest c/a ratio (1.886) \cite{Kittel} (c is
the length between equivalent\ basal planes and a is the distance between
nearest neighbors within the hexagonal structure) and is thus far removed from
the ideal hcp crystal $\surd$8/3 (=1.633) which refers to the closest possible
stacking of spheres in a hexagonal lattice. This fact might explain why in
these chemically synthesized nanocrystals only very flat, two dimensional
crystals are formed.

Besides characterization by transmission electron microscropy (TEM), x-ray
diffraction patterns on 12 different samples (not shown here) revealed a
strong preferred orientation along the (002) direction parallel to the
underlying glass substrate used for the measurement. This is consistent with
earlier measurements on evaporative thin films of cadmium.\cite{Halder}\ In
the majority of x-ray diffraction results for these chemically produced
cadmium nanocrystals, the (002) peak was the only peak in the diffraction
spectra. It is important to note that these crystals were not susceptible to
oxidation (no CdO diffraction peaks were observed in the x-rays results and
there was no indication of the reddish tint of CdO\ in solution) which
supports our hypothesis that the nonpolar portion of oleic acid is combining
with the cadmium nanocrystals.%

\begin{figure}
[ptb]
\begin{center}
\includegraphics[
natheight=3.010400in,
natwidth=3.000000in,
height=3.0536in,
width=3.0441in
]%
{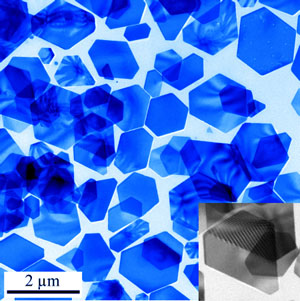}%
\caption{(Color Online) TEM\ image of equilateral and non-equilateral cadmium
hexagons produced by reduction of cadmium acetate. The sizes of these hexagons
range from 150 nm to 700 nm in diameter. \textit{Inset}: Moire patterns are
observed between two overlaping hexagons.}%
\end{center}
\end{figure}

Figure 2 shows an TEM image of a single hexagonal spiral. While the segments
of the spiral are well delineated and stable, there is also a large rounded
body of mass at the end of the spiral. Also in the image there are clear
patterns of dark and light colored fringes indicating the bending modes as the
crystal forms and overlaps with itself. It is noteworthy that in the lower
right of the crystal, the emerging bend is evident as dark black lines extend
from a central region as the crystal begins to turn and form another segment
of its six fold geometry. The separation of the segments could be due to oleic
acid or simply empty space. As the nanocrystals were cleaned in various
polar/nonpolar solvents before placing them in distilled toluene, the effect
of cleaning may have removed the interior oleic acid which is what we believe
keeps the separation between the segments of the hexagon when forming in
solution. This configuration was the only one of its kind observed in
TEM\ images produced from this particular reaction (although only a small
portion of the reaction product was sampled).%
\begin{figure}
[ptb]
\begin{center}
\includegraphics[
natheight=3.000000in,
natwidth=3.000000in,
height=3.0441in,
width=3.0441in
]%
{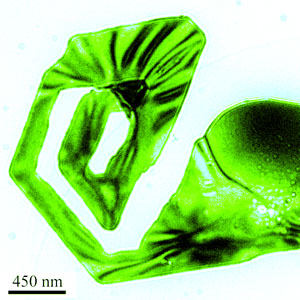}%
\caption{(Color Online) TEM image of a single spiraling hexagon made in
solution.}%
\end{center}
\end{figure}

Another reaction produced additional spiraling hexagons under slightly
different conditions. In Figure 3, TEM images of 9 individual spiraling
hexagons are shown. Again it should be emphasized that these were not the
dominant configurations found in solution and were selected amongst other
shapes which were not spiraling hexagons. In the TEM images of Figure 3, it is
observed that there is a feeder attached to several spiraling hexagons. This
asymmetric feeder appears to be responsible for the overall growth and
resulting size of the nanocrystals. Although the nature of the feeder is still
unclear, it is likely that it contains large numbers of loose two dimensional
Cd nuclei that are free to move inside the spiral. It also appears to be
responsible for the outermost edge of the spiral being perfectly straight,
whereas the interior segments of some spirals seems to deviate from an ideal
hexagonal geometry. Since the spirals are grown from inside out, it is
possible that the interior slightly looses its posture as the crystals are
swirling around in solution before the growth process is quenched.
\begin{figure}
[ptb]
\begin{center}
\includegraphics[
natheight=3.010400in,
natwidth=3.000000in,
height=3.0536in,
width=3.0441in
]%
{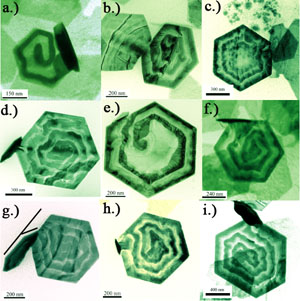}%
\caption{(Color online) TEM images of nine different spiraling hexagons taken
from the same solution. In Image\textit{ g}, lines are drawn as an aid to the
eye to indicate how the "feeder" is aligned with the nanocrystal. Also in
Image\textit{ g }there are bending modes which extend vertically across the
hexagon.}%
\end{center}
\end{figure}

The feeder of Fig. 3i is enlarged in Fig. 4. In this expanded image, the
feeder appears to be three dimensional with structure akin to an asymmetric
hexagon.
\begin{figure}
[ptb]
\begin{center}
\includegraphics[
natheight=3.000000in,
natwidth=3.000000in,
height=3.0441in,
width=3.0441in
]%
{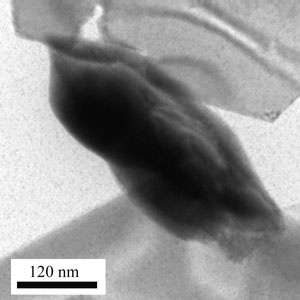}%
\caption{Enlarged image of the "feeder" shown in Fig. 3i believed to be
responsible for the spiraling hexagons. The shape, with its tapered edges, is
suggestive of an asymmetric hexagon.}%
\end{center}
\end{figure}

There are notable exceptions to this apparent feeder as shown in the
TEM\ images of Fig. 3 b \& c. Here the spiraling hexagons are apparently
feeding off non-spiraling hexagons. Since these are well developed spirals the
dense feeder may have been present at an earlier stage of its growth.

The rounded mass feature that was a part of the spiraling hexagon shown in
Fig. 2 is also evident in the images of Fig.3 a \& e. These rounded features
do not appear to be mechanically or artificially attached but rather appear to
be an integral part of the hexagons like the one shown in Fig. 2. In Fig. 3e,
the starting point, or head of the spiraling hexagon, is itself a hexagon. The
rounded feature follows and is not at the end of the spiraling hexagon like
the previous cases. The growth mechanism of this particular hexagon also does
not appear to rely on a feeder or adjoining hexagon. However, it may be the
case that the feeder was removed in the process of preparing the sample for TEM\ imaging.

The dimensionality of the spiraling hexagons is assumed to be two dimensional.
This assumption is based on two key features observed in the TEM\ images of
Figure 3. \ One is the electron transparency through spiraling hexagons, and
the other is the appearance of bending modes that can span the breadth of the
hexagons ( see for example Fig. 3g).%
\begin{figure}
[ptb]
\begin{center}
\includegraphics[
natheight=3.000000in,
natwidth=3.000000in,
height=3.0441in,
width=3.0441in
]%
{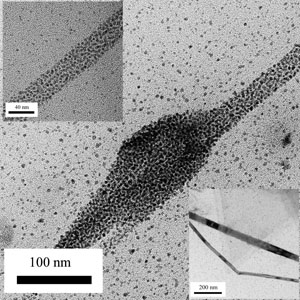}%
\caption{TEM\ image of an early stage of growth as nuclei form a nanowire with
a rounded protrusion. \textit{Top Inset}: TEM image of a segment of the same
nanowire. \textit{Bottom Inset}: TEM\ image of nanowires (from a different
reaction) that do not spiral.}%
\end{center}
\end{figure}

Is the distinctiveness of the spirals a consequence of thermal agitation by
the electron beam used when imaging these structures? In Figure 5, TEM\ images
of straight nanowires that do not spiral under identical imaging conditions
are presented as counterexamples to spiraling hexagons. These cadmium
nanowires were synthesized from a modified version of the same reaction that
produced spirals and demonstrate that the beam of electrons from the TEM is
not triggering the spiraling effect. The main physical difference between
spiraling and nonspiraling hexagons is the presence of the feeder and
chemically the main difference is the higher concentration of oleic acid used
in the synthesis of the spiraling hexagons. The rounded protrusion seen in
Figure 5 is likely a consequence of an early stage growth of the hexagonally
shaped end that is a standard feature in the crystalline nanowires.

The possibility that this is a manifestation of pattern formation is now
addressed. Experimentally the study of pattern formations in chemical
reactions is most lucid in Belousov-Zhabotinsky (BZ) reactions \cite{Bel}
where spiral patterns spontaneously emerge.\cite{Jak} The present situation is
different from the Turing instability\cite{Tur} in that the spiraling pattern
has six-fold symmetry. In addition much of the work on pattern formation is
built around surface tension driven effects arising from a nondeformable
liquid-gas interface. In most cases these effects involve a temperature
gradient. This work addresses surface tension effects from liquid - solid
interfaces using a colliodial technique in which there is a density gradient.

In summary, we presented experimental evidence for the formation of growth
spirals with six-fold symmetry without the assistance of screw or edge
dislocations from a substrate. These growth spirals were produced chemically
and appear to be two dimensional. Spiraling in plane might be indicative of
the presence of only two dimensional nanocrystals, the absence of a substrate,
cadmium's unusually large c/a ratio \cite{Kittel} and the increased
concentration of oleic acid used in the growth process. The main physical
feature relevant to these spiraling hexagons is a feeder, either in the form
of an aggregate of nanocrystals attached to the spiral or a neighboring
hexagon. The position of the attached feeder could be interpreted as the same
position that an analogous screw dislocation would have, however in this study
it appears to be dynamical.

Future directions. Atomic force microscope analysis of the samples to obtain a
height profile is necessary to confirm the dimensionality of these spirals and
map height differences across a spiraling hexagon.\ Also these spiraling
hexagons could possibly be exploited as a new type of metamaterial due to
their unique configuration. Finally a detailed theoretical description needs
to be established to address the features that are observed here.

Acknowledgements. The author thanks Heinrich Jaeger, Horst Strunk, John Royer,
Robert Josephs,William Sweeney and Allen Goldman. This work was supported by
the University of Chicago-Argonne Consortium for Nanoscience Research. The use
of shared experimental facilities provided by the Chicago MRSEC is also
gratefully acknowledged.

*Current address: The Physics Department and Center of Nanophysics and
Advanced Materials at the University of Maryland, College Park. lladams@umd.edu

\bigskip

\end{document}